# Transient heat transfer analysis of fins with variable surface area and temperature-dependent thermal conductivity using an integral transform technique

Marcos Filardy Curi[1,*], José Luiz Zanon Zotin[1]

[1]Centro Federal de Educação Tecnológica Celso Suckow da Fonseca, CEFET/RJ – UnED Itaguaí, Mechanical Engineering Department

Rodovia Mário Covas, Lote J2, Quadra J, Itaguaí, Rio de Janeiro, P.O. Box 23812-101, Brazil; Tel.: +55 21 2688 3570

Marcos Filardy Curi ORCID: https://orcid.org/0000-0001-8216-3564

José Luiz Zanon Zotin ORCID: https://orcid.org/0000-0003-3289-374X

[*]Corresponding author: marcos.curi@cefet-rj.br; jose.zotin@cefet-rj.br

# ABSTRACT

This paper aims at the transient heat transfer analysis on extended surfaces with temperature-dependent thermal conductivity, constant internal heat generation, and five different geometries. The governing equations developed in this work consider the effects of the function that describes the shapes studied. To ensure a more effective thermal analysis, we investigated the impact of the fin's surface area and its arc length on the convection term, and the influence of the overall function of the fin profile affecting the volumetric rate of stored thermal energy. The imposed boundary conditions are adiabatic type on the tip of the fin and prescribed base temperature. A hybrid mathematical method, known as the Generalized Integral Transform Technique (GITT), was used to solve the cases presented here. The results obtained with GITT were firstly validated with FEM results, presenting excellent agreement between them, proving been a method suitable for handling non-linear problems. Physical effects on the temperature distribution and efficiency due to the thermo-geometric parameter (M), internal heat generation (Q), and variable thermal conductivity were investigated. The analysis shows that increasing M or decreasing Q results in a longer transient period, regardless of the geometry assessed. Besides, the efficiency of the increasing linear shape is higher than those of the rectangular fin for any parameter choice. Moreover, depending on the thermo-geometric value, the rectangular fin becomes more efficient among the remaining shapes. However, for higher values, those differences become undistinguished from each other, demonstrating a significant impact of the varying cross-sectional area and the fin surface area on the overall thermal analysis.

Nomenclature

| | |
|---|---|
| $A_c$ | Fin cross-sectional area |
| $A_{cb}$ | Fin base cross-sectional area |
| $A_s$ | Fin surface area |
| $Bi$ | Biot Number |
| $C_p$ | Heat capacity |
| $f$ | Fin profile function |
| $F$ | Dimensionless fin profile function |
| $g$ | Arc length differential |
| $G$ | Dimensionless arc length differential |
| $h$ | Convective heat transfer coefficient |
| $k$ | Thermal conductivity |
| $k_0$ | Thermal conductivity at $T_\infty$ |
| $L$ | Length of the fin |
| $M$ | Thermo-geometric parameter |
| $n$ | Fin geometric parameter |
| $N_i$ | Normalization integral of the eigenfunction |
| $N_T$ | Truncation order |
| $\dot{q}$ | Internal heat generation |
| $Q$ | Dimensionless internal heat generation |
| $Q_b$ | Heat diffused through the fin base |
| $Q_f$ | Fin heat loss |
| $Q_I$ | Ideal heat loss |
| $t$ | Time |
| $s$ | Arc length |
| $T$ | Temperature |
| $T_b$ | Fin base temperature |
| $T_\infty$ | Surrounding fluid temperature |
| $x$ | Spatial coordinate |

| | | |
|---|---|---|
| | $X$ | Dimensionless spatial coordinate |
| | $W$ | Width of the fin |
| Greek Symbols | | |
| | $\beta$ | Dimensionless thermal conductivity gradient |
| | $\delta$ | Fin thickness at the base |
| | $\delta_0$ | Difference between the fin base and the fin end thickness |
| | $\eta$ | Fin efficiency |
| | $\theta$ | Dimensionless temperature |
| | $\theta_b$ | Fin base dimensionless temperature |
| | $\lambda$ | Parameter controlling the thermal conductivity variation |
| | $\mu$ | Eigenvalues |
| | $\tau$ | Dimensionless time |
| | $\varphi$ | Geometric Parameter |
| | $\psi$ | Eigenfunctions |
| | $\omega$ | Geometric Parameter |
| Superscripts and subscripts | | |
| | $*$ | Filtered quantity |
| | $^-$ | Transformed quantity |
| | $\sim$ | Normalized quantity |
| | $i, j, k$ | Indices of the eigenfunction $\psi$ |

# 1. INTRODUCTION

Heat transfer problems in mechanical engineering have a growing number of studies that focus its efforts on how to increase the efficiency of specific devices for heat dissipation. The analysis of extended surfaces could be mentioned as a specific application for this purpose. Some areas are directly affected, such as heat dissipation for computer processors, refrigeration of fluids passing by ducts in a thermodynamic system, in the automotive and aerospace sectors, among others. The study shows fundamental importance not only in the optimization of energy consumption but also cutting costs by reducing the equipment sizes. Therefore, the fin geometry could be optimized to maximize the efficiency of heat dissipation. The literature has seen a large number of recent studies on heat transfer provided by extended surfaces, contributing to valuable information and numerical data. An extensive review of this topic is done by Kraus et al. [1].

Trying to capture a closer behavior with real heat transfer phenomena in extended surfaces, one first assumption widely found in literature is the effect on thermal analysis due to variable properties, e.g., the thermal conductivity. For a large temperature gradient in the fins, the thermal conductivity becomes a function of the temperature profile turning it to a non-linear problem. Convection heat transfer coefficient, heat generation, and surface emissivity could also be some examples of temperature dependence [2-3]. The important investigation on how this temperature dependence affects the study of longitudinal fins have been carried out by a great number of theoretical analyses, where many researchers used different mathematical tools to solve the non-linear nature of the problem, then, some valuable studies in this sense are introduced.

Joneidi et al. [4] used the semi-analytical method known as Differential Transformation Method (DTM) and Patra and Ray [5] used the Homotopy Perturbation Sumudu Transform Method (HPSTM), both for solving the temperature field and efficiency of convective straight fins with temperature-dependent thermal conductivity. They also analyzed the physical effects of thermo-geometric fin parameter and thermal conductivity parameters on the heat transfer. Coşkun and Atay [6], performed this same analysis with the Variational Iteration Method (VIM), and to show the reliable results from VIM, made a comparison with available results from Adomian Decomposition Method (ADM) and the Finite Element Methods (FEM). Ghasemi et al. [7] used the Differential Transformation Method (DTM) for solving the convective straight fins with temperature dependence for internal heat generation and thermal conductivity, revealing the effectiveness of the employed method. Sobamowo [8], using the finite difference method, showed that the fin temperature distribution and the fin efficiency are significantly affected by the thermo-geometric parameters of the fin. Later, Sobamowo [9], adds the effect of the variable convective heat transfer coefficient. With no heat generation, Anbarloei and Shivanian [10] studied the problem with temperature-dependent thermal conductivity and variable convective coefficient, obtaining exact analytical expressions for temperature field and efficiency as a function of the thermo-geometric fin parameter.

Huang and Li [11] and Ganji et al. [12], using an integral transform method and DTM, respectively, studied a similar approach without heat generation for a stationary convective-radiative problem, obtaining an expression for the temperature field and efficiency. The effects of thermo-geometric fin parameter and thermal conductivity parameters on the temperature distribution were also analyzed. Dogonchi and Ganji [13] included a temperature-dependent heat generation term and performed an analysis for a fin with a longitudinal rectangular profile using DTM, showing that the tip temperature of the fin increases for higher values for heat generation. Singh et al. [14] studied a highly non-linear problem for convective-radiative fins, where the thermal conductivity, convective heat transfer coefficient, and surface emissivity are all temperature-dependent; the last one also varies with the wavelength of the material according to the black body theory.

The studies above are based on thermal analyses for fin problems with variable properties, steady-state conditions, and constant cross-section area, suitable for many applications. However, it is possible to find a growing number of research that studied the transient heat transfer in extended surfaces in the last 10 years. Besides that, a key factor contributing to overall thermal analysis is the study for different fin profiles that indicate the best shape to maximize the fin efficiency. In that sense, it is worth mentioning the following works.

Khan and Aziz [15] studied the transient temperature field of a convective-radiative problem for longitudinal straight fins with constant internal heat generation and constant properties. They concluded that the fin acts as a heat sink instead of a heat source if the heat generation exceeds a threshold. Mhlongo et al. [16] analyze the transient behavior of longitudinal rectangular fins, with variable thermal conductivity and heat transfer coefficient. The use of symmetry techniques shows the effect on temperature profile due to the thermo-geometric fin parameter and the variable properties. Ma et al. [17] used the Spectral Collocation Method (SCM) to solve the transient heat transfer problem of convective-radiative fins, with variable thermal conductivity, heat generation, convective heat transfer coefficient, and emissivity. The boundary conditions are convection type on the tip of the fin. They show accurate results for the transient temperature and efficiency under the influence of all involved parameters. Vyas et al. [18] studied the transient for the convective-radiative formulation, constant proprieties, and different fin profiles using the Meshless Local Petrov Galerkin Method (MLPG), the temperature-time history has shown closer values to real problems.

Mosayebidorcheh et al. [19] also performed a transient analysis of extended surfaces with variable fin profiles and temperature-dependent for thermal conductivity, convection heat transfer coefficient, and heat generation. They used a hybrid technique based on DTM and Finite Difference Method (FDM). It is possible to note the effect of the fin profile shape on the temperature distribution and efficiency, as well the effects of variable properties. Later, Mosayebidorcheh et al. [20], considering a stationary state, included the radiation terms as a function of temperature for different fin profiles, solving it through the Least Square Method (LSM). The results have shown the direct dependence on the heat transfer rate across the fin with its thickness ratio. Ganji and Dogonchi [21], studying different fin shapes with the variable cross-sectional area, used DTM for longitudinal fins problem with temperature dependence for internal heat generation and thermal conductivity, analyzing the temperature inside the fin for different values of the parameters.

In the work of Turkyilmazoglu [22], exact solutions for a straight extended surface with exponential shape was carried out, with thermal conductivity and convective heat transfer coefficient as a function of the temperature. Analyses of the fin best shape is investigated for different conditions, showing that the efficiency of exponential profiles is higher than those for the constant cross-sectional area. Also, Turkyilmazoglu [23] adds for discussion the influence of fins in movement and with internal heat generation, stating once more the advantages of using exponential profiles. Roy et al. [24] studied the effects on the temperature profile with a simultaneous variation of thermal conductivity, convective heat transfer coefficient, surface emissivity, and internal heat generation, for three different geometries. The ADM and the Modified Adomian Method (MADM) were used to solve the non-linear formulation. On the other hand, Torabi and Zhang [25], used the DTM for the thermal analysis of convective–radiative from longitudinal fins with different geometries. They reported the effects of the physically applicable parameters on the temperature profiles and efficiency.

As introduced, the studies carried out the influence of variable properties, transient behavior, and different geometries on the extended surface thermal analysis. All employed some diverse mathematical methods such as DTM, ADM, FDM, among others, providing reliable results and agreement with numerical data previously published. In this work, a hybrid mathematical method, known as the Generalized Integral Transform Technique (GITT) was used to solve the cases presented here. Widely used in different classes of problems, including diffusive and convective-diffusive non-linear problems [26-28]. It is an expansion of the classic method of integral transform technique (CITT) [26], introducing an inherently straightforward hybrid analytical-numerical solution, based in eigenfunctions expansion from a Sturm-Liouville problem that contains information from the original partial differential equation [29-33]. As mentioned above, many works in the literature use GITT to solve heat transfer problems and show its effectiveness, however with rare contributions for extended surface problems.

This paper aims to assess how each geometric and physical parameter impact on the temperature distribution along the fin and its global efficiency. Thereby, we investigate the transient heat transfer problem on fins with temperature-dependent thermal conductivity and different geometries. In this work, the proposed formulation of the governing equations considers the influence of the extended surface profile function on the transient term, which directly affects the volumetric rate of stored thermal energy. Also, the influence of the fin surface area working on the external convection for each geometry is considered. This approach differs from the classical and widely used formulation for extended surface, that neglects the simultaneous contribution of these factors, even in the works which focus their attention on transient analyses with different fin shapes.

## 2. PROBLEM FORMULATION

The present study aims to solve the transient heat transfer problem on a fin with variable geometry along its length in addition to internal heat generation and temperature-dependent thermal conductivity. Consequently, as explained above, the consideration of a more realistic problem means to increase the non-linearity of the governing partial differential equation, requiring an increasingly robust solution method.

## 2.1 Fin Geometry

To evaluate the geometric influence on the overall fin thermal analysis, this study considered different geometries for one-dimensional longitudinal fins, represented by:

$$f(x) = \begin{cases} \dfrac{(\delta - \delta_0)}{2} + \dfrac{\delta_0 L^n}{2(L-x)^n}, & n \leq 0 \\ \dfrac{\delta}{2} + \dfrac{\delta_0}{2}\left(\dfrac{x}{L}\right)^n, & n > 0 \end{cases} \quad (1)$$

where $\delta$ is the fin thickness at the base, $\delta_0$ is the difference between the fin base and the fin end thickness, and $L$ and $W$ are the length and width of the fin, respectively, as noted in Fig. 1. It is also important to notice that Eq. (1) presents two different functions used to create those geometries, with the thickness increasing or decreasing along the fin length depending on the exponent $n$ and its signal. So, in this work, five different geometries were considered: one rectangular ($n = 0$), two trapezoidal ($n = 1$ and $n = -1$), and two convex parabolic ($n = 2$ and $n = -2$).

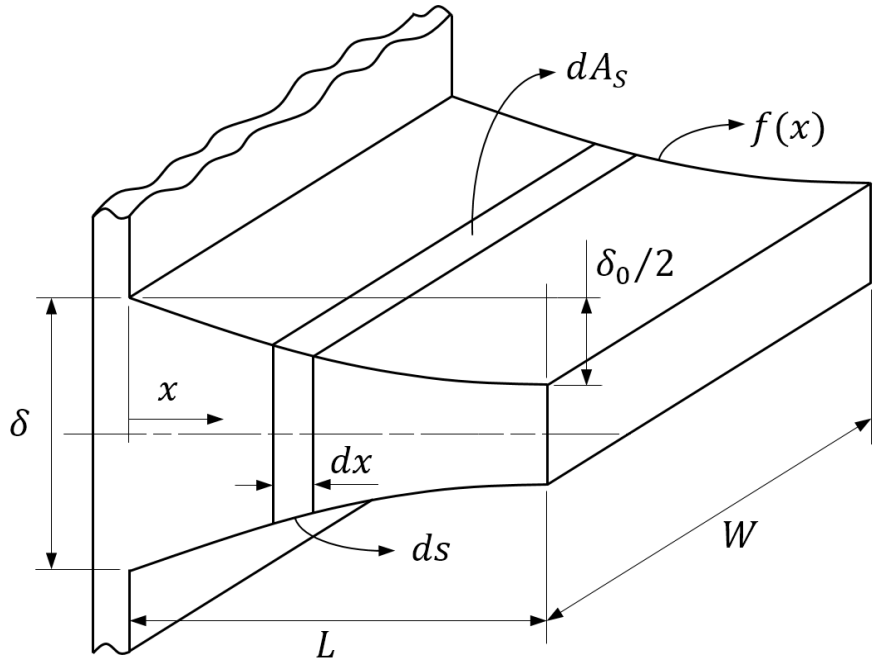

**Figure 1: Longitudinal fin profile.**

## 2.2 Fin Heat Transfer Problem

Often, problems involving extended surfaces are non-linear. For some ordinary thermal analyses, they are transformed into linear ones by many assumptions, such as fin shapes and constant properties, widely mentioned earlier. However, for large temperature gradients, the involved thermophysical properties become a function of the temperature profile, turning it to a non-linear problem. For this reason, many works have developed the theory that surrounds heat transfer across extended surfaces, which adapt the classic heat diffusion approach to a simplified case where heat transfer occurs in only one direction, due to the geometric characteristics of the fins.

So, extending the formulation proposed by Kraus et al. [1], the transient formulation for the energy balance with temperature-dependent thermal conductivity for a longitudinal fin with general geometry, illustrated in Fig.1, and with the assumption that the fin thickness is negligible as compared to its width, $W \gg \delta$, can be expressed as:

$$\rho C_p \frac{\partial T(x,t)}{\partial t} A_c(x)dx$$
$$= \frac{\partial}{\partial x}\left(A_c(x)k(T)\frac{\partial T(x,t)}{\partial x}\right)dx - h(T(x,t) - T_\infty)dA_s \quad (2a)$$
$$+ \dot{q}A_c(x)dx$$

$$k(T) = k_0\big(1 + \lambda(T(x,t) - T_\infty)\big) \quad (2b)$$

$$A_c(x) = 2Wf(x); \quad A_s = 2Ws = 2W\int_0^L \sqrt{1 + f'(x)^2}dx \; ; \quad (2c,d)$$

$$dA_s = 2W\sqrt{1 + f'(x)^2}dx = 2Wg(x)dx \quad (2e)$$

where $T_\infty$, $h$ and $\dot{q}$ are the surrounding fluid temperature, convective heat transfer coefficient, and internal heat generation, respectively. Also, $\rho$ and $C_p$ are the fin material density and heat capacity, respectively. The variable thermal conductivity adopted from [34] is given by Eq. (2b), where $k_0$ is defined as the thermal conductivity at $T_\infty$, and $\lambda$ is the parameter controlling the thermal conductivity variation. The formulation proposed in this work shows the effects of the variable cross-sectional area ($A_c$) and the fin surface area ($A_s$), due to variable geometry along the x-direction, noted in Eq. (2a). In Eq. (2c) and (2d) these areas are defined as a function of the fin profile, $f(x)$, and its arc length, $s$.

On the right-hand side of Eq. (2a) we could note the influence of the fin geometry on terms inherent from internal phenomena, which depends on the amount of material limited by $f(x)$, i.e., the diffusion and internal heat generation terms. An important contribution, barely seen in literature, is the fin profile acting in the convection term through the external surface area variation, where the arc length differential is represented by $g(x)$, which is a function of $f(x)$. On the left-hand side, we also consider the effects of $f(x)$ on the transient term. This influence must be pointed out since this term is expressed by the volumetric rate of stored thermal energy within the fin. All these considerations, in the same analysis, is not observed in the literature of extended surface. So, applying Eq. (2b,c,e) into Eq. (2a) one can obtain the following heat transfer equation:

$$\rho C_p f(x)\frac{\partial T(x,t)}{\partial t} = \frac{\partial}{\partial x}\left(f(x)k(T)\frac{\partial T(x,t)}{\partial x}\right)$$
$$-h(T(x,t) - T_\infty)g(x) + \dot{q}f(x), \quad 0 \leq x \leq L, \; t \geq 0 \quad (3a)$$

along with the boundary conditions adopted:

$$T(0, t) = T_b; \qquad \left.\frac{\partial T(x,t)}{\partial x}\right|_{x=L} = 0; \qquad t \geq 0 \qquad (3b,c)$$

$$T(x, 0) = T_\infty, \qquad 0 \leq x \leq L \qquad (3d)$$

where $T_b$ is the fin base temperature. As can be noticed, it was considered a Dirichlet boundary condition at $x = 0$, with an insulated fin tip. Defining the following dimensionless groups:

$$\theta = \frac{T(x,t) - T_\infty}{T_b - T_\infty}; \quad X = \frac{x}{L}; \quad \tau = \frac{t\,\alpha}{L^2}; \quad \beta = \lambda(T_b - T_\infty); \quad \omega = \frac{\delta_0}{\delta};$$
$$Bi = \frac{hL}{k_0}; \quad \varphi = \frac{\delta}{L}; \quad M^2 = \frac{Bi}{\varphi}; \quad Q = \frac{L^2 \dot{q}}{k_0(T_b - T_\infty)} \qquad (4)$$

where $\alpha$ is the thermal diffusivity expressed as $\alpha = k_0/\rho C_p$, M is the thermo-geometric parameter, $\beta$ is the dimensionless thermal conductivity gradient, and $Q$ is the dimensionless internal heat generation. In this article, the geometric parameters $\omega$ and $\varphi$ were fixed as $\omega = 0.8$ and $\varphi = 1/3$. Applying Eq. (4) in Eq. (3) it is possible to obtain the following dimensionless formulation of this problem:

$$F(X)\frac{\partial \theta(X,\tau)}{\partial \tau} = F(X)\left(\beta\left(\frac{\partial \theta(X,\tau)}{\partial X}\right)^2 + (1 + \beta\theta(X,\tau))\frac{\partial^2 \theta(X,\tau)}{\partial X^2}\right)$$
$$+ (1 + \beta\theta(X,\tau))\frac{\partial F(X)}{\partial X}\frac{\partial \theta(X,\tau)}{\partial X} - \frac{Bi}{\varphi}\theta(X,\tau)G(X) + QF(X) \qquad (5a)$$
$$, \ 0 \leq X \leq 1, \ \tau \geq 0$$

$$\theta(0,\tau) = \theta_b = 1; \qquad \left.\frac{\partial \theta(X,\tau)}{\partial X}\right|_{X=1} = 0; \qquad \tau \geq 0 \qquad (5b,c)$$

$$\theta(X, 0) = 0, \qquad 0 \leq X \leq 1 \qquad (5e)$$

where $F(X)$ is the fin dimensionless geometric profile, depicted in Fig. 2, and $G(X)$ the dimensionless derivative of the fin arc length, given by:

$$F(X) = \frac{f(x)}{\delta} = \begin{cases} \frac{(1-\omega)}{2} + \frac{\omega}{2(1-X)^n}, & n \leq 0 \\ \frac{1}{2} + \frac{\omega X^n}{2}, & n > 0 \end{cases} \qquad (5f)$$

$$G(X) = \sqrt{1 + \varphi^2 F'(X)^2} \qquad (5g)$$

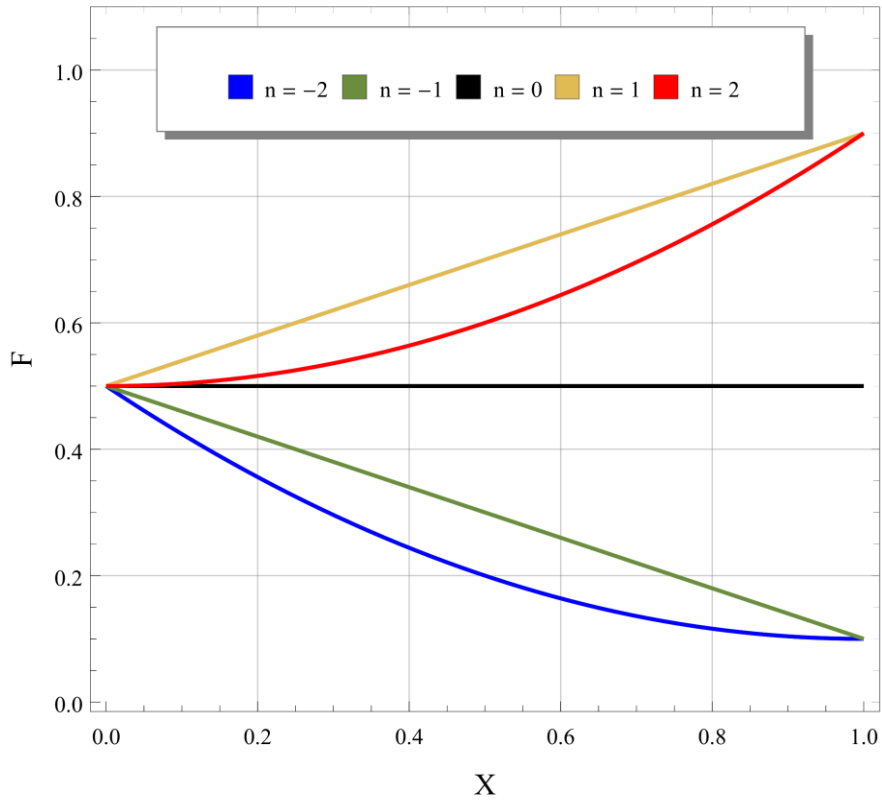

**Figure 2: The different dimensionless fin geometries adopted**

2.3 Solution Methodology

The method chosen to solve Eq. 5 was the Generalized Integral Transform Technique (GITT), a hybrid numerical-analytical method often used in convective-diffusive problems. Through an integral transform pair and a Sturm-Liouville eigenvalue problem, it is possible to transform the partial differential equation proposed into a coupled system of ordinaries differential equations (ODEs).

This system does not have an analytical solution, requiring a numerical approach to solve it. Then, the potential solution can be obtained using the inverse function of the transformed pair, resulting in an eigenfunction expansion that must be truncated to achieve the required solution convergence [26-33]. To improve the computational performance of the present solution, and reduce the numbers of terms in the eigenfunction expansion, it is recommended to homogenize the boundary conditions using a simple filter defined as:

$$\theta(X,\tau) = \theta^*(X,\tau) + 1 \qquad (6)$$

where $\theta^*$ is the filtered potential. Applying Eq. (6) into Eq. (4), the following filtered problem can be obtained:

$$F(X)\frac{\partial\theta^*(X,\tau)}{\partial\tau} = F(X)\left(\beta\left(\frac{\partial\theta^*(X,\tau)}{\partial X}\right)^2 + (1+\beta+\beta\theta^*(X,\tau))\frac{\partial^2\theta^*(X,\tau)}{\partial X^2}\right)$$
$$+(1+\beta+\beta\theta^*(X,\tau))\frac{\partial F(X)}{\partial X}\frac{\partial\theta^*(X,\tau)}{\partial X} \quad (7a)$$
$$-\left(\frac{Bi}{\varphi}G(X) + \frac{Bi}{\varphi}\theta^*(X,\tau)G(X)\right) + QF(X)$$
$$, 0 \leq X \leq 1, \quad \tau \geq 0$$

$$\theta^*(0,\tau) = 0; \qquad \left.\frac{\partial\theta^*(X,\tau)}{\partial X}\right|_{X=1} = 0; \qquad \tau \geq 0 \qquad (7b,c)$$

$$\theta^*(X,0) = -1, \qquad 0 \leq X \leq 1 \qquad (7e)$$

With the filtered problem properly defined, it is now possible to apply the GITT formalism through a transform/inverse pair given by:

$$\text{Transform:} \quad \bar{\theta}_i^*(\tau) = \int_0^1 \theta^*(X,\tau)\,\tilde{\psi}_i(X)dX \qquad (8a)$$

$$\text{Inverse:} \quad \theta^*(X,\tau) = \sum_i^\infty \tilde{\psi}_i(X)\bar{\theta}_i^*(\tau) \qquad (8b)$$

where $\tilde{\psi}_i(X)$ are the normalized eigenfunctions, expressed as:

$$\tilde{\psi}_i(X) = \frac{\psi_i(X)}{\sqrt{N_i}} \qquad (8c)$$

$$N_i = \int_0^1 \psi_i^2(X)dX \qquad (8d)$$

The eigenfunctions $\psi_i(X)$ can readily be achieved through the Sturm-Liouville problem solution, the simplest eigenvalue problem that can be proposed, and its boundary conditions given by:

$$\frac{\partial^2\psi_i(X)}{\partial X^2} + \mu_i^2\psi_i(X) = 0 \qquad (9a)$$

$$\psi_i(0) = 0, \qquad \left.\frac{\partial\psi_i(X)}{\partial X}\right|_{X=1} = 0 \qquad (9b,c)$$

The solution of Eq. (9a) for different boundary conditions can be found in Ozisik [35]. Finally, applying the operator $\int_0^1 (\cdot)\,\tilde{\psi}_i(X)dX$ on Eq. (7) and making use of the inverse formulae in Eq. (8b), it is possible to obtain the transformed problem and the transformed initial condition, given by the following system of ODEs:

$$\sum_{j=1}^{\infty} \frac{\partial \bar{\theta}_j^*(\tau)}{\partial \tau} D_{i,j} = \sum_{k,j=1}^{\infty} \bar{\theta}_j^*(\tau) \bar{\theta}_k^*(\tau) A_{i,j,k} + \sum_{j=1}^{\infty} \bar{\theta}_j^*(\tau) B_{i,j} + C_i \quad (10a)$$

$$\bar{\theta}_i^*(0) = -\int_0^1 \tilde{\psi}_i(X) dX \quad (10b)$$

where:

$$A_{i,j,k} = \beta \left( \int_0^1 F(X) \frac{\partial \tilde{\psi}_j(X)}{\partial X} \frac{\partial \tilde{\psi}_k(X)}{\partial X} \tilde{\psi}_i(X) dX \right.$$
$$- \mu_k^2 \int_0^1 F(X) \tilde{\psi}_k(X) \tilde{\psi}_j(X) \tilde{\psi}_i(X) dX \quad (10c)$$
$$\left. + \int_0^1 \frac{\partial F(X)}{\partial X} \frac{\partial \tilde{\psi}_k(X)}{\partial X} \tilde{\psi}_j \tilde{\psi}_i(X) dX \right)$$

$$B_{i,j} = \left( (1+\beta) \left( \int_0^1 \frac{\partial F(X)}{\partial X} \frac{\partial \tilde{\psi}_j(X)}{\partial X} \tilde{\psi}_i(X) dX - \mu_j^2 \int_0^1 F(X) \tilde{\psi}_j(X) \tilde{\psi}_i(X) dX \right) \right.$$
$$\left. - \frac{Bi}{\varphi} \int_0^1 G(X) \tilde{\psi}_j \tilde{\psi}_i(X) dX \right) \quad (10d)$$

$$C_i = -\frac{Bi}{\varphi} \int_0^1 \tilde{\psi}_i(X) G(X) dX + Q \int_0^1 \tilde{\psi}_i(X) F(X) dX \quad (10e)$$

$$D_{i,j} = \int_0^1 F(X) \tilde{\psi}_i(X) \tilde{\psi}_j(X) dX \quad (10f)$$

The system defined in Eq. (10) does not have an analytical solution, being necessary a numerical approach to solve it after its truncation to an order $N_T$. In this case, it was chosen the *NDSolve* routine from *Wolfram Mathematica* software [36], which solves non-linear ODEs systems with absolute and relative error controls predetermined. With the numerical solution obtained, it is possible to recover the dimensionless temperature field $\theta(X,\tau)$ by applying the inverse formula into Eq. (6), resulting in an approximated numerical-analytical expansion solution given by:

$$\theta(X,\tau) = 1 + \sum_i^{N_T} \tilde{\psi}_i(X) \bar{\theta}_i^*(\tau) \quad (11)$$

It should be noted that the truncation order $N_T$ must be chosen to achieve the desired solution convergence.

## 2.4 Fin Efficiency

To verify how the geometry and each parameter influence the fin thermal exchange, it is necessary to calculate, for each case, the fin efficiency under steady-state condition, which can be defined as the ratio of the actual heat loss ($Q_f$) to the ideal heat loss ($Q_I$). The actual heat loss is the heat dissipated by the fin to the surrounding environment, which is the heat diffused through the fin base ($Q_b$). On the other hand, the ideal heat loss is calculated considering all the fin surfaces kept at the fin base temperature, which would guarantee the maximum heat flux possible for a specific fin. So, the actual and ideal heat loss can be expressed as:

$$Q_f = Q_b = -k(T)A_{cb} \frac{dT}{dx}\bigg|_{x=0} \tag{12a}$$

$$Q_I = hA_s(T_b - T_\infty) \tag{12b}$$

where $A_{cb}$ is the fin base cross-section area and $A_s$ is the entire fin surface area. Consequently, the fin efficiency $\eta$ can be obtained by employing the dimensionless groups from Eq. (4) into Eq. (12), yielding:

$$\eta = -\frac{(1+\beta\theta_b)\frac{d\theta}{dX}\bigg|_{X=0}}{2\frac{Bi}{\varphi}\theta_b \int_0^1 \sqrt{1+\varphi^2 F'(X)^2}\,dX} \tag{13}$$

## 3. RESULTS AND DISCUSSION

A preliminary analysis between the GITT employed in this work with other numerical methods found in the literature was performed to ensure its feasibility and reliability for the obtained results. In this case, the method chosen was the FEM, through the Finite Element solution method from the NDSolve routine of the Wolfram Mathematica v. 12 software [36], where it was specified the precision and accuracy number of 6 digits. Figure 3 shows comparisons of the temperature profile for all fin geometries. To maintain consistency in the analysis and the method validation, we considered fixed values for the main dimensionless groups found in Eq. 4. Thus, 3 results are presented for each $n$, with the combinations of $\beta = 0.5, Bi = 1.5, Q = 0; \beta = -0.5, Bi = 1, Q = 0$ and $\beta = 0.5, Bi = 1, Q = 1$ in $\tau = 2$.

It is possible to notice that the comparison between GITT and FEM shows excellent consistency between them, which can be confirmed through the numerical analysis seen in Table 1 and Table 2. Both were made for time instants $\tau = 0.1$, in the transient period, $\tau = 2$, in the stationary regime and $\beta = -0.5, Bi = 1, Q = 2, n = -2$ and $\beta = 0.5, Bi = 1, Q = 1, n = 2$, respectively. Thus, it is noted that even for problems with indices $n = \pm 2$, making the problem with a high degree of non-linearity, the cases show a numerical agreement in the 5th significant digit. Another relevant characteristic is that we need only 50 terms to truncate the sum and achieve an excellent convergence of values for the temperature field, described by Eq. (11). In this way, it is possible to state that the adopted model can provide solutions with excellent accuracy for the problem proposed in this article.

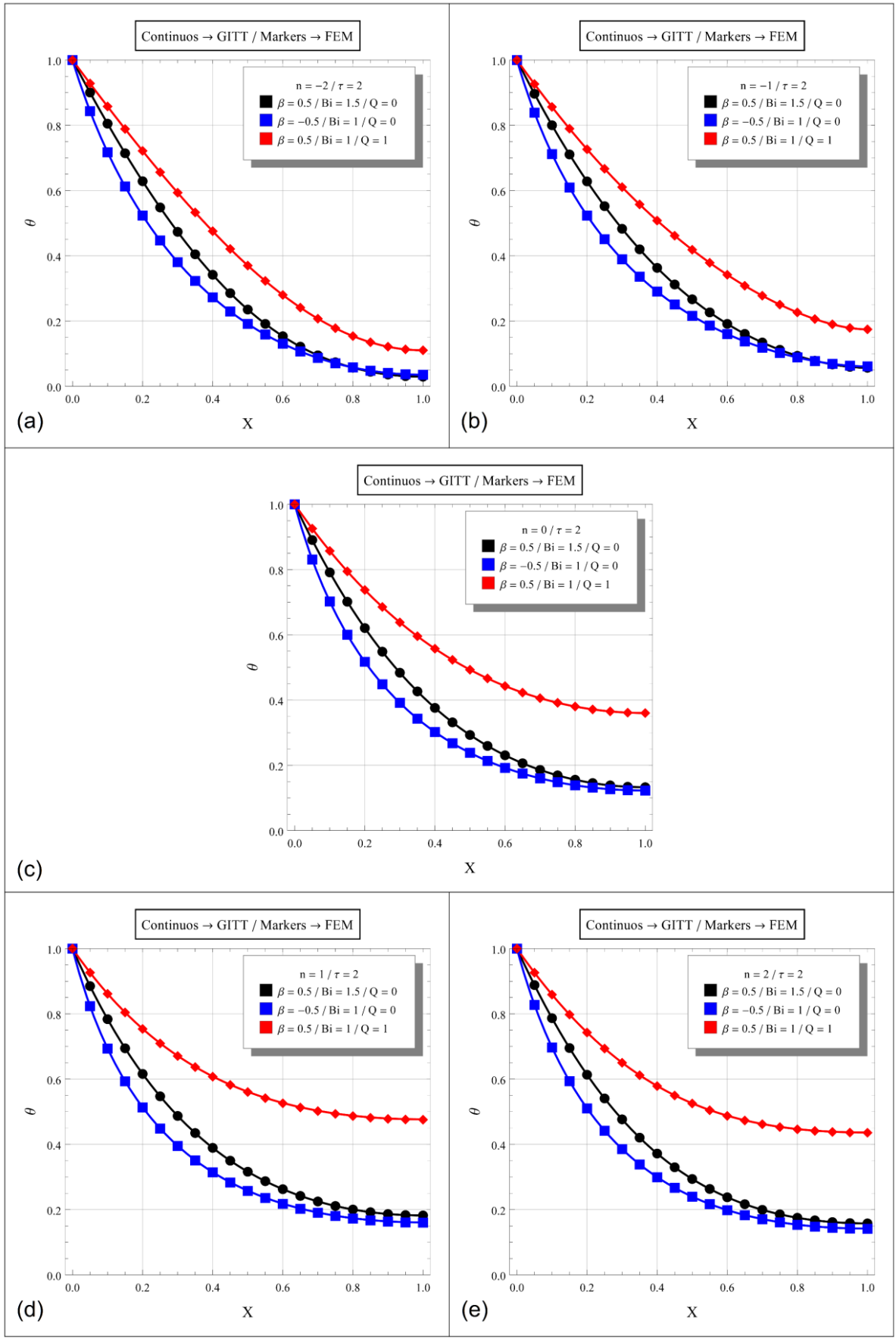

**Figure 3: Comparison between GITT and FEM results.**

**Table 1: Dimensionless temperature convergence behavior and comparison with FEM for $n = -2$.**

| $n = -2$ | $\beta = -0.5 \ / \ Bi = 1 \ / \ Q = 2$ | | | | | | | |
|---|---|---|---|---|---|---|---|---|
| | $\tau = 0.1$ | | | | $\tau = 2$ | | | |
| | X=0.25 | X=0.5 | X=0.75 | X=1 | X=0.25 | X=0.5 | X=0.75 | X=1 |
| $N = 10$ | 0.49679 | 0.24383 | 0.12240 | 0.08636 | 0.54145 | 0.29347 | 0.15702 | 0.11183 |
| $N = 20$ | 0.49567 | 0.24372 | 0.12278 | 0.08679 | 0.54043 | 0.29338 | 0.15737 | 0.11221 |
| $N = 30$ | 0.49572 | 0.24373 | 0.12275 | 0.08683 | 0.54047 | 0.29338 | 0.15734 | 0.11225 |
| $N = 40$ | 0.49576 | 0.24373 | 0.12273 | 0.08684 | 0.54051 | 0.29338 | 0.15733 | 0.11226 |
| $N = 50$ | 0.49576 | 0.24373 | 0.12274 | 0.08684 | 0.54051 | 0.29338 | 0.15733 | 0.11226 |
| FEM | 0.49575 | 0.24373 | 0.12274 | 0.08684 | 0.54051 | 0.29338 | 0.15733 | 0.11227 |

**Table 2: Dimensionless temperature convergence behavior and comparison with FEM for $n = 2$.**

| $n = 2$ | $\beta = 0.5 \ / \ Bi = 1 \ / \ Q = 1$ | | | | | | | |
|---|---|---|---|---|---|---|---|---|
| | $\tau = 0.1$ | | | | $\tau = 2$ | | | |
| | X=0.25 | X=0.5 | X=0.75 | X=1 | X=0.25 | X=0.5 | X=0.75 | X=1 |
| $N = 10$ | 0.57315 | 0.29544 | 0.15856 | 0.12200 | 0.69412 | 0.52579 | 0.45345 | 0.43631 |
| $N = 20$ | 0.57294 | 0.29542 | 0.15863 | 0.12209 | 0.69387 | 0.52577 | 0.45353 | 0.43642 |
| $N = 30$ | 0.57295 | 0.29542 | 0.15862 | 0.12210 | 0.69388 | 0.52577 | 0.45353 | 0.43643 |
| $N = 40$ | 0.57296 | 0.29542 | 0.15862 | 0.12210 | 0.69389 | 0.52577 | 0.45352 | 0.43643 |
| $N = 50$ | 0.57296 | 0.29542 | 0.15862 | 0.12210 | 0.69389 | 0.52577 | 0.45352 | 0.43643 |
| FEM | 0.57295 | 0.29541 | 0.15862 | 0.12211 | 0.69388 | 0.52577 | 0.45352 | 0.43644 |

We begin the thermal analysis and its implications from Figure 4. It shows the temperature profiles for all extended surface geometries obtained for $Bi = 1$ and $Q = 0$, with the parameter $\beta$ varying between 0.5, 0, and $-0.5$, from top to bottom, respectively. The figures on the left - Fig.4 (a), (c), and (e) - presents the temperature distribution at two different locations ($X = 0.5$ and $X = 1$) over time, emphasizing the transient behavior. At $\tau = 2$, all fin geometries considered achieved a fully stationary regime. The figures on the right - Fig.4 (b), (d) and (f) - shows the temperature field along the fin at two dimensionless time instants, $\tau = 0.05$ and $\tau = 2$. Each instant chosen refers to the transient period and the stationary regime, respectively.

In the stationary regime, one can notice that a decrease in $\beta$ implies a more significant temperature difference between the fin base (maximum temperature) and the fin tip (minimum temperature) and vice versa. For $\beta$ positive, it is possible to observe that temperatures remain at higher values along the entire fin than progressively lower $\beta$'s. At the initial transient regime, these temperature profiles are superimposed between the different fin geometries. And only with long periods of time, it is possible to notice the effect of heat transfer phenomena on the fin temperature distribution. For progressively smaller $\beta$ a longer time is necessary to reach the steady state. Since $\beta$ is a parameter linked to thermal conductivity, positive values imply an increase in the material conductivity, as can be noticed in Eq. 2b. Therefore, the thermal energy diffuses more efficiently, keeping the temperature at higher values, clearly implying into fin efficiency, whose analysis will be discussed later.

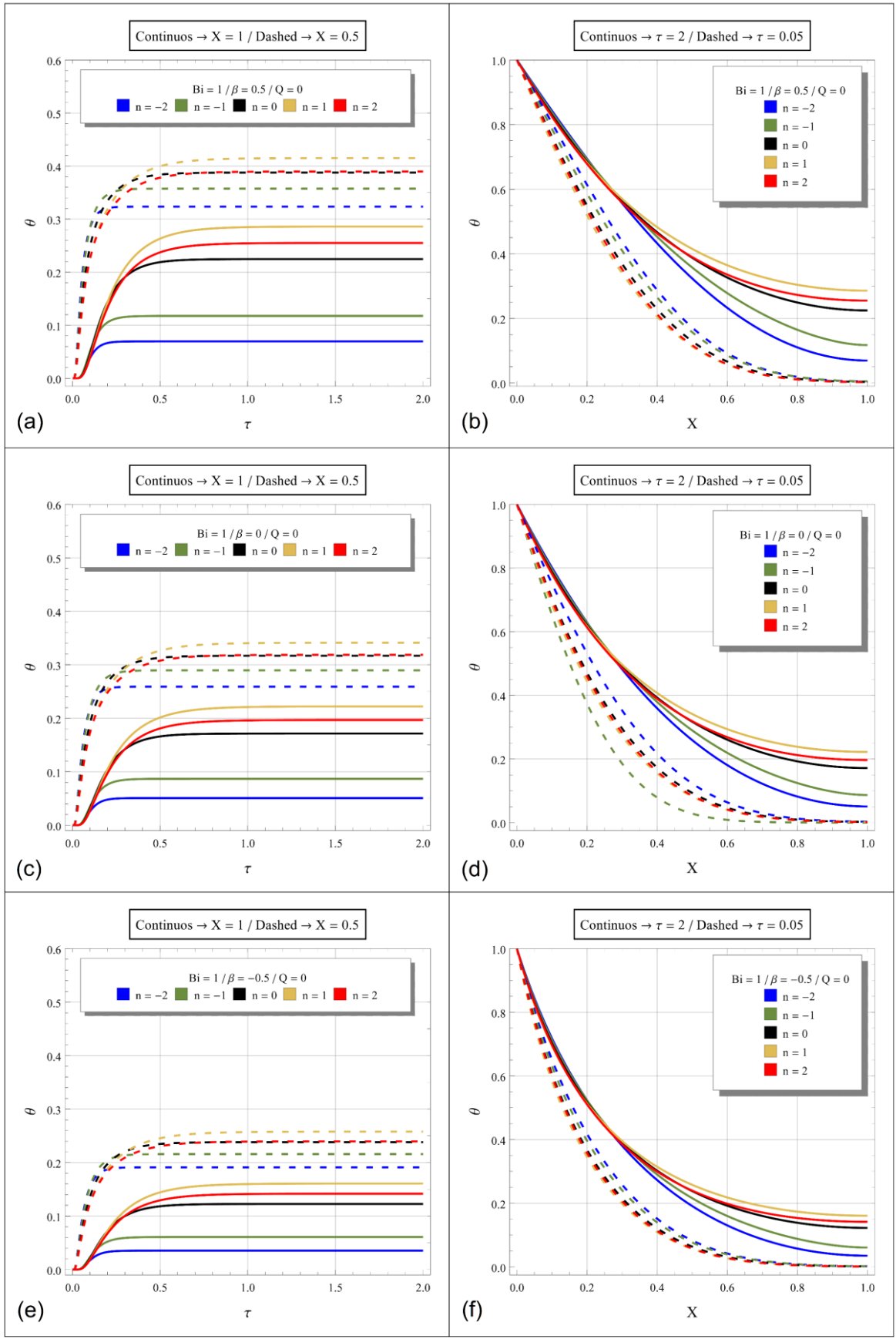

**Figure 4: Temperature profiles within different fins geometries for different $\beta$ values.**
(a) and (b) $\beta = 0.5$, (c) and (d) $\beta = 0$, (e) and (f) $\beta = -0.5$.

When decreasing $\beta$, Figures 4(b), (d), and (f), the temperature profiles for each $n$ tends to approach between them, reaching lower values in the direction of $X = 0$. This behavior can also be verified in the transient results from Figures 4(a), (c), and (e). Extrapolating this analysis, a progressive decrease in $\beta$ with values even less than $-0.5$ would result in temperature profiles practically overlapping at any point of the fin, regardless of the time instant and the fin geometry. This consideration would result in minimum temperatures closer to the base ($X = 0$), characterizing the use of materials with low conductivity, not ideal in fins applications.

Also, in Fig. 4, it is possible to verify that in stationary cases, the shape of the trapezoidal fin ($n = 1$) maintains the temperature in higher levels along the entire fin when compared with the other geometries. In the case of $n = -2$, the opposite occurs. This divergence becomes even more evident over the fin length, especially from $X = 0.5$. These results indicate that the fin geometry, and consequently, the amount of material that constitutes it, directly influences the temperature distribution. With more material, it is possible to accumulate a greater amount of thermal energy. Thus, by ordering the temperature distribution along the fin in decreasing order, one obtains $n = 1, 2, 0, -1, -2$. This conclusion corroborates with Figs. 4 (a), (c), and (e), which represent the temporal evolution of temperature. It is possible to notice that for $n = 1$, a longer time is necessary to reach the steady-state condition, while for $n = -2$, a much shorter time is spent. In other words, the function $F(X)$, which influences the transient term seen in Eq. (3a) and determines the amount of fin material, shows that the thermal energy saturation throughout the fin takes more time following the decreasing order $n = 1, 2, 0, -1, -2$.

Figure 5 shows the influence of the fin thermo-geometric parameter ($M$), presented in Eq. 4. With the formulation proposed in this article, we have an alternative way of representing it, defining $M$ as a function of $Bi$ and $\varphi$, without losing the original physical meaning. Because $\varphi = 1/3$ was considered in all simulations, any $Bi$ variation presented can be considered directly proportional to $M$. For $Bi = 1$, one can observe, for all different geometries, a difference in temperature along the fin, between the base and the tip, smaller when compared to higher $Bi$ values. This phenomenon is perceptible in the stationary regime, where the temperature distribution remains at higher values for low $Bi$ along the entire fin length. Also, when increasing Bi, the temperature reaches lower values toward $X = 0$. For example, we can see from Figures 5 (b), (d) and (f) that in $\tau = 2$ and for n = -1, we have $\theta = 0.2$ in approximately $X = 0.8$, $X = 0.6$ and $X = 0.5$ for $Bi = 1$, $Bi = 1.5$ and $Bi = 2$, respectively. In the transient regime, it is possible to note similar results, although, in initial instants, the temperature distributions for different $F(X)$ are almost overlapping.

These physical behaviors analyzed through $Bi$ are directly associated with thermal conductivity or the convective coefficient, as seen in Eq. 4. With greater conductivity, the heat diffusion is more efficient through the fin, resulting in a temperature profile with higher values in the entire fin. The use of low conductivity materials, which would return high $Bi$ values to the problem, would not be ideal in a real fin application. However, a higher $Bi$ can also be associated with higher convective coefficient $h$, which means a more intense external convection, lowering all temperature profiles.

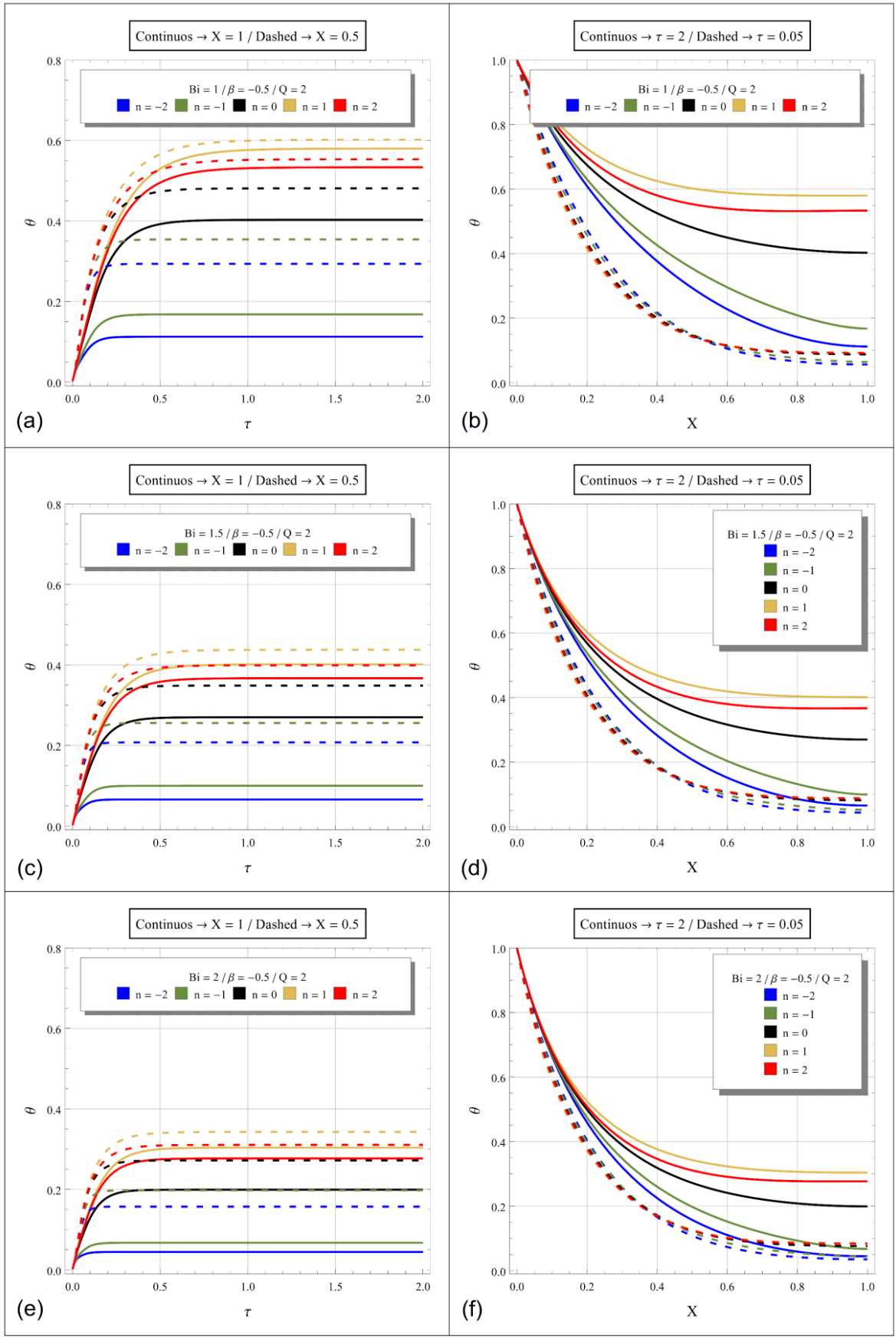

**Figure 5: Temperature profiles within different fins geometries for different $Bi$ values.**
(a) and (b) $Bi = 1$, (c) and (d) $Bi = 1.5$, (e) and (f) $Bi = 2$.

Fig. 6 demonstrates the variation of the internal heat generation ($Q$) and its impacts on the fin's different shapes, for $\beta = 0.5$ and $Bi = 1.5$. Again, the temperature distribution as a function of position for $\tau = 0.05$ and $\tau = 2$ and as a function of time for $X = 0.5$ and $X = 1$, are arranged in the same way as Figs. 4 and 5. The values adopted for $Q$ were 0, 1, and 2, from top to bottom, respectively. We can notice the considerable change in the temperature distribution with the increase of $Q$. For progressively higher $Q$, the temperature distribution assumes higher values for any shape along the entire fin, which can be seen more clearly in Figures 6 (b), (d), and (f). Naturally, as it is an energy generation parameter, its increase will necessarily cause the internal temperature values to rise. Thus, the transient behavior, seen in Figures 6 (a), (c), and (e), are also directly impacted, requiring more time to saturate and reach the steady state for higher values of $Q$. Fig. 6 complements the whole analysis inherent to the heat transfer of the proposed problem carried out so far with the variations of the parameters $\beta$ and $Bi$.

Figures 7, 8, and 9 illustrate the effects of parameters $\beta$, $Bi$, and $\varphi$, respectively, on the heat transfer efficiency along the fin in the stationary regime for the geometries contemplated in this article; rectangular ($n = 0$), trapezoidal ($n = 1$ e $n = -1$) and convex parabolic ($n = 2$ and $n = -2$). In Figure 7, we analyze the efficiency varying $\beta$ for $Bi = 1.5$, $Q = 0$, and $\varphi = 1/3$, where it is observed that the efficiency increases with the increase of $\beta$. From Eq. 13 the efficiency behavior in Fig. 7 is expected since the efficiency $\eta$ is directly proportional to $\beta$. For higher $\beta$ values, the temperature distribution (Fig. 4) along the entire fin also shows high values, resulting in a greater temperature difference with the ambient fluid $T_\infty$, and consequently, in better efficiencies for any geometry. The trapezoidal geometry, $n = 1$, has better efficiency results among the others. However, the efficiency for $n = 2$ and $n = 0$ presents very similar values, since the temperature profiles (Fig. 4) for these two geometries were almost overlapping in both transient and permanent analysis, with the fixed parameters adopted. It is possible to state that the better the thermal conductivity, the better the thermal diffusion along the extended surface, and the heat transfer becomes more efficient.

The effect of the thermo-geometric parameter ($M$) on fins efficiency is seen in Fig. 8, with the number of Biot varying between 0 and 10 and considering as fixed parameters $\beta = 0.5$, $Q = 0$, and $\varphi = 1/3$. Similar behavior of decrease in the fin efficiency is observed for all geometries with the increase of $Bi$, which is an expected behavior after analyzing the temperature distribution along the fin made in Fig. 5. For values between $Bi = 0$ and $Bi = 3$, an abrupt drop in efficiency is observed. From $Bi = 5$, this drop drastically decreases, presenting a behavior with a slight discrete variation in efficiency, regardless of the geometry. We also observed that for $Bi$ tending to zero, efficiency tends to 1, that is, 100%, which would correspond to the analysis of a case with the fin thermal conductivity tending to infinite (negligible thermal resistance), and with the fin reaching the maximum temperature of the problem along its entire surface. If the convective heat transfer coefficient ($h$) is null, meaning that the environment does not remove heat from the fin, or $L = 0$ (without a fin), one can arrive at the same previous conclusion. However, these considerations would represent an unreal or purposeless case of study in the case of fins. Inversely proportional to $Bi$, this effect is predicted by Eq. 13 and directly implies the efficiency values achieved. The Biot number, in addition to being represented by the ratio between the convective heat transfer coefficient and the thermal conductivity, also represents the ratio between the conduction and convective heat transfer resistances.

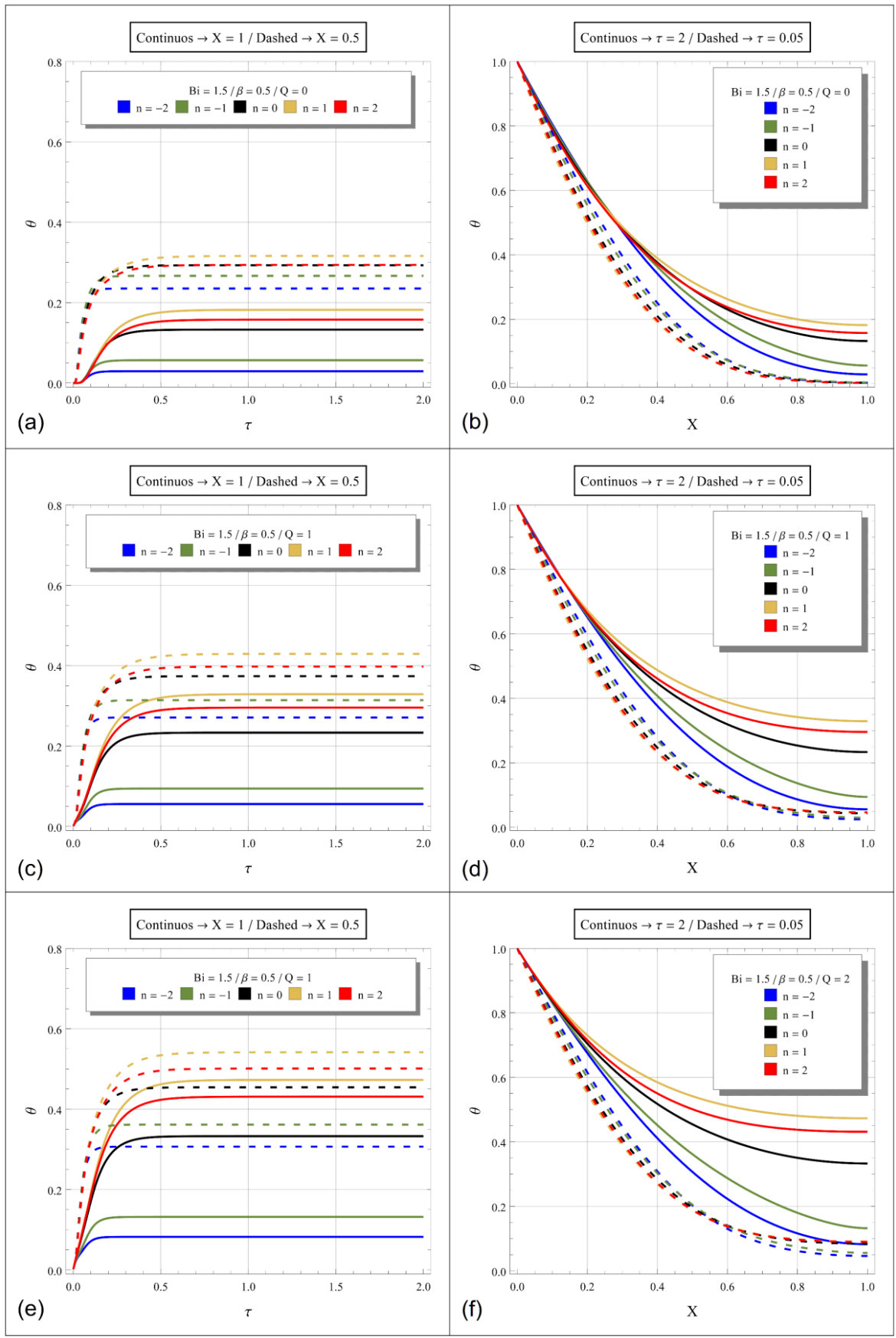

**Figure 6: Temperature profiles within different fins geometries for different $Q$ values.**
(a) and (b) $Q = 0$, (c) and (d) $Q = 1$, (e) and (f) $Q = 2$.

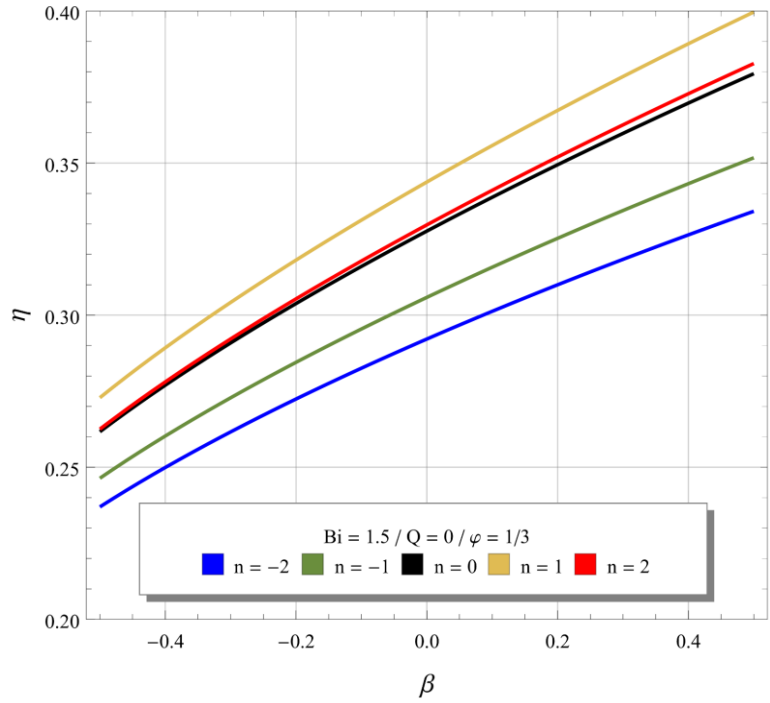

**Figure 7: Fin efficiency for different $\beta$ values and different geometries.**

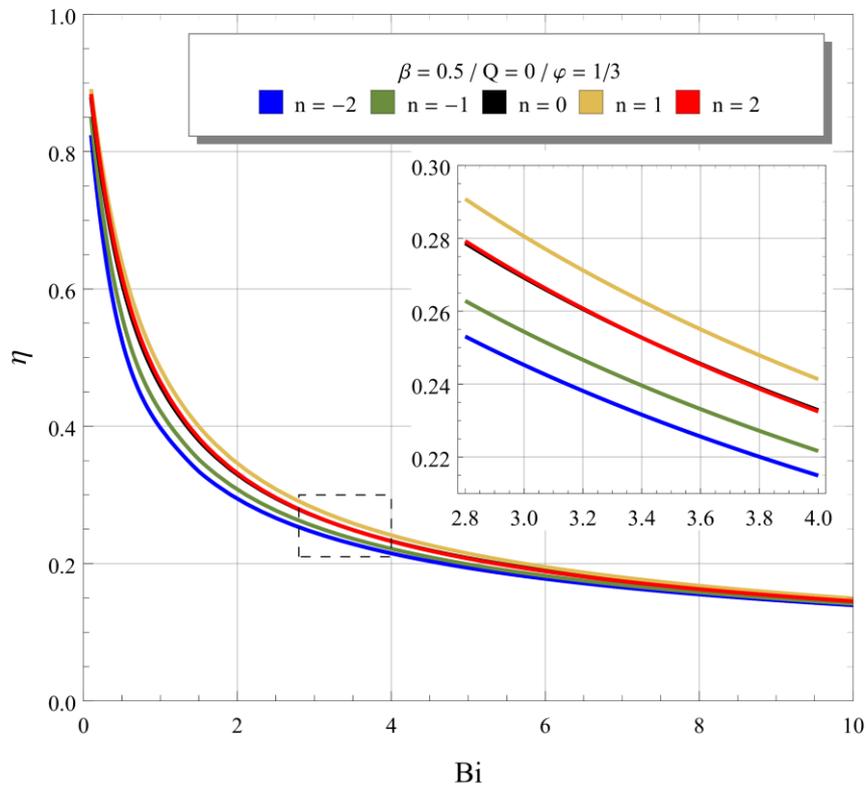

**Figure 8: Fin efficiency for different $Bi$ values and different geometries.**

Therefore, increasing $Bi$, we have a greater conduction heat transfer resistance along the fin, making it difficult to transport thermal energy, lowering its efficiency. On the other hand, if we increase the fin length ($L$), we will have more material that constitutes it, increasing the conduction heat transfer resistance and lowering the efficiency. One can also observe from Eq. 13 that $L$ is directly related to the different geometries expressed by $F(X)$, which is associated with the surface area $dA_s$ (Eq. 2 (d)), inversely proportional to $\eta$. Thus, if $L$ tends to infinity, the surface area in contact with the environment will also tend to infinity, resulting in a fin efficiency tending to zero, since a large part of the fin will be at room temperature $T_\infty$ (only positions close to $X = 0$ will show some difference). In this case, the distinction between the fin's shapes would be negligible for a higher range of $Bi$, e.g., very long lengths, obtaining lower and practically identical values for efficiency. These results states that the intrinsic non-linearity of the problem vanishes in that range. It is possible to conclude that the fin length (not too long) and the material it is made of (good conductivity), are essential to determine the best configuration that returns the maximum efficiency for a specific application. Fig. 8 also presents a zoomed-in region for a better distinction between each geometry's results, where it is possible to observe a better efficiency for trapezoidal geometries ($n = 1$), regardless of $Bi$. For geometries $n = 0$ and $n = 2$, the efficiencies are again very similar, with an inversion in $Bi = 3.4$.

The analysis of Fig. 9 provides information on the behavior of efficiency as a function of $\varphi$, considering $\beta = 0.5$, $Q = 0$, and $Bi = 1.5$. We observed a better efficiency for $n = 1$ and $n = -2$ the worst efficiency, regardless of $\varphi$, with an inversion between $n = 0$ and $n = 2$ for $\varphi = 0.5$. For high values of $\varphi$, where $L$ is low, or $\delta$ is high, we will have a short fin compared to its thickness, limiting the area in contact with the ambient fluid and the amount of material contained by $F(X)$. As a result, there is a decrease in thermal resistance by conduction, increasing efficiency. For smaller values of $\varphi$, we have a long and thin configuration, lowering efficiency regardless of the geometry considered.

Therefore, we noted that no geometry with the fin tip thinner than its base (i.e., $n < 0$) can be more efficient than an ordinary rectangular fin. In the case of an $n = 2$ parabolic fin, depending on its geometry and the problem's thermal parameters, it can have a greater or lesser efficiency compared to a rectangular fin. However, the efficiency gain would not be greater than 1%, making it less attractive than the rectangular fin. In the case of a trapezoidal geometry $n = 1$, all the analyses made in this article indicated superior efficiency to all the other analyzed geometries. Nonetheless, it also has a larger volume among all these geometries, which means that its production cost would be higher. In this case, a cost-benefit analysis of this type of fin is essential to assess its feasibility concerning an easy-to-build common rectangular fin.

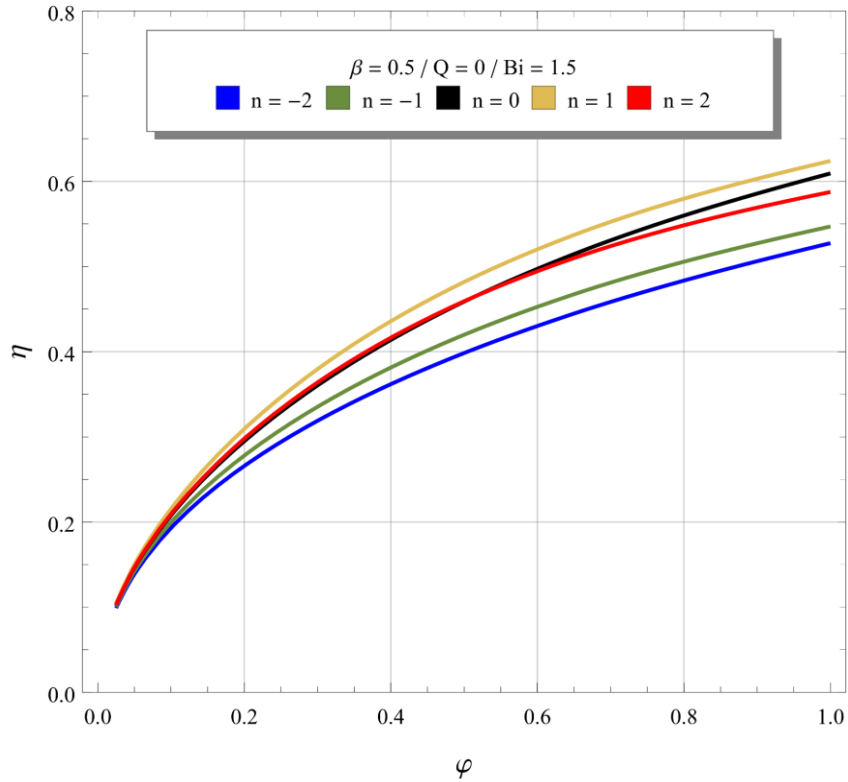

**Figure 99: Fin efficiency for different $\varphi$ values and different geometries.**

## 4. CONCLUSIONS

A one-dimensional formulation for a transient heat transfer problem in fins with an adiabatic tip and different types of geometry is presented in this article, including heat generation and temperature-dependent thermal conductivity. The formulation includes the surface area $A_s$ as a function of the fin profile $f(x)$ and its arc length $s$, responsible for evaluating the heat transfer loss more accurately through the fin surface area. Besides that, the transient term of the PDE also includes $f(x)$ to properly represent the volumetric rate of stored thermal energy within the fin. The methodology presented here to solve these problems was the Generalized Integral Transform Technique (GITT), whose results were verified through the Finite Element Method (FEM). The comparison presented an excellent agreement, guaranteeing the developed code validation.

We presented several results for the transient and steady state temperature profile for each geometry considered and varying the dimensionless parameters $\beta$, $Bi$, and $Q$, representing the effect of the thermal conductivity, the external convection, and the internal heat generation, respectively. Thereby, it was possible to verify that:

- increasing $\beta$, i.e., increasing the growth rate of the material thermal conductivity, will also increase the temperature distribution along the fin since the heat will be better diffused from the base to the fin tip. Consequently, higher fin efficiencies can be achieved.

- increasing $Bi$, i.e., increasing the external convection or reducing the material thermal conductivity, will decrease the temperature distribution along the fin since a greater conduction heat transfer resistance exists. Consequently, the fin efficiency will decrease. A lower thermal conductivity will also lead to a lower diffusivity, increasing the time necessary to achieve the steady-state condition.
- increasing $Q$, will increase the temperature distribution along the fin since more thermal energy is added to the system. Besides that, this extra thermal energy saturates the fin material more rapidly, lowering the transient period.
- in all cases verified, the trapezoidal $n = 1$ geometry was the most efficient one, and the geometry with $n < 0$ was the less efficient. The rectangular $n = 0$ and the parabolic $n = 2$ presented very similar efficiencies, showing some variations depending on the values of $\beta$, $Bi$, and $\varphi$.
- decreasing $\varphi$, i.e., increasing the fin length or decreasing its thickness, will decrease the fin efficiency since the conduction heat transfer resistance is increased.

## 5. ACKNOWLEDGEMENTS

The authors are grateful for all support offered by the Centro Federal de Educação Tecnológica Celso Suckow da Fonseca, CEFET/RJ – Campus Itaguaí.